\documentclass[11pt, logo, nonumbering]{preprint}
\usepackage[utf8]{inputenc}
\usepackage{graphicx}
\usepackage{amsmath}
\usepackage{booktabs}
\usepackage{xcolor}
\definecolor{mydarkblue}{rgb}{0,0.08,0.45}
\usepackage[colorlinks=true,linkcolor=mydarkblue,citecolor=mydarkblue,filecolor=mydarkblue,urlcolor=mydarkblue]{hyperref}
\usepackage{xspace}
\usepackage{cleveref}
\usepackage{enumitem}
\usepackage{multirow}
\usepackage{pifont}
\newcommand{\cmark}{\ding{51}}
\newcommand{\xmark}{\ding{55}}
\usepackage{tabularx}

\usepackage[style=numeric-comp,maxbibnames=3,minnames=1,backend=bibtex, doi=false,eprint=false,url=false]{biblatex}

\definecolor{gred}{RGB}{250, 210, 207}
\definecolor{coolblue1}{rgb}{0.91, 0.94, 0.98}
\definecolor{coolblue2}{rgb}{0.76, 0.85, 0.94}
\definecolor{coolblue3}{rgb}{0.54, 0.72, 0.87}
\definecolor{coolblue4}{rgb}{1, 1, 1}
\definecolor{codegreen}{rgb}{0,0.6,0}
\definecolor{codegray}{rgb}{0.5,0.5,0.5}
\definecolor{codepurple}{rgb}{0.58,0,0.82}
\definecolor{backcolour}{rgb}{0.97,0.97,0.97}

\usepackage{listings}
\lstdefinestyle{mystyle}{
    backgroundcolor=\color{backcolour},
    commentstyle=\color{codegreen},
    keywordstyle=\color{magenta},
    numberstyle=\tiny\color{codegray},
    stringstyle=\color{codepurple},
    basicstyle=\ttfamily\small,
    breakatwhitespace=false,
    breaklines=true,
    captionpos=b,
    keepspaces=true,
    numbers=left,
    numbersep=5pt,
    showspaces=false,
    showstringspaces=false,
    showtabs=false,
    tabsize=2,
    frame=single,
    framerule=0.5pt
}
\begin{document}

\title{Building Enterprise Realtime Voice Agents from Scratch:  \\ A Technical Tutorial}

\author{
Jielin Qiu, Zixiang Chen, Liangwei Yang, Ming Zhu, Zhiwei Liu, Juntao Tan, \\ 
Wenting Zhao, Rithesh Murthy, Roshan Ram, Akshara Prabhakar, \\
Shelby Heinecke, Caiming, Xiong, Silvio Savarese, Huan Wang \\
~~~~\\
\textsuperscript{}Salesforce AI Research \\
~~~~\\
\href{https://github.com/SalesforceAIResearch/enterprise-realtime-voice-agent}{
  \raisebox{-0.3\height}{\includegraphics[height=0.8cm]{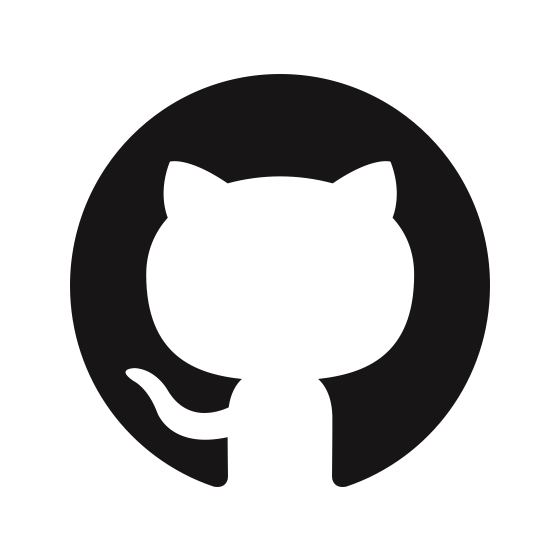}}
  \textbf{https://github.com/SalesforceAIResearch/enterprise-realtime-voice-agent}
}
}

\maketitle

\begin{abstract}
We present a technical tutorial for building enterprise-grade realtime voice agents from first principles. While end-to-end speech-to-speech models may ultimately provide the best latency for voice agents, fully self-hosted end-to-end solutions are not yet available. We evaluate the closest candidate, Qwen3-Omni, across three configurations: its cloud-only DashScope Realtime API achieves $\sim$702ms audio-to-audio latency with streaming, but is not self-hostable; its local vLLM deployment supports only the Thinker (text generation from audio, 516ms), not the Talker (audio synthesis); and its local Transformers deployment runs the full pipeline but at $\sim$146s---far too slow for realtime. The cascaded streaming pipeline (STT $\rightarrow$ LLM $\rightarrow$ TTS) therefore remains the practical architecture for self-hosted realtime voice agents, and the focus of this tutorial. We build a complete voice agent using Deepgram (streaming STT), vLLM-served LLMs with function calling (streaming text generation), and ElevenLabs (streaming TTS), achieving a measured time-to-first-audio of 755ms (best case 729ms) with full function calling support. We release the full codebase as a 9-chapter progressive tutorial with working, tested code for every component.
\end{abstract}

\section{Introduction}
\label{sec:intro}

The emergence of large language models (LLMs) with strong reasoning and tool-use capabilities has created demand for voice-based AI agents that can converse naturally while executing complex tasks, i.e., booking appointments, querying databases, managing orders. While text-based agents are well-understood, adding a real-time voice interface introduces fundamental engineering challenges around latency, streaming, and audio processing that are poorly documented in existing literature.

\paragraph{The landscape is fragmented.} Our survey reveals a striking gap between available components and integrated systems:

\begin{itemize}[nosep]
    \item \textbf{25+ open-source speech-to-speech models} exist (Moshi~\cite{moshi2024}, Qwen3-Omni~\cite{qwen3omni2025}, Qwen2.5-Omni~\cite{qwen25omni2025}, GLM-4-Voice~\cite{glm4voice2024}, Kimi-Audio~\cite{kimiaudio2025}, Step-Audio~\cite{stepaudio2025}, LLaMA-Omni~\cite{llamaomni2024}, Mini-Omni~\cite{miniomni2024}, Freeze-Omni~\cite{freezeomni2024}, among others~\cite{wavchat2024,spiritlm2024}), but none ships with a tutorial on building a realtime voice agent.
    \item \textbf{Production frameworks} like Pipecat~\cite{pipecat2024} and LiveKit Agents~\cite{livekitagents2024} provide excellent plumbing, but are ``pip install and use'' tools that do not teach \textit{how} they work internally.
    \item \textbf{Evaluation frameworks} like Benchforce~\cite{benchforce2024} define enterprise environments (hospital, HR, order management) with function calling, but use turn-based (non-streaming) voice pipelines.
    \item \textbf{No codebase} builds a complete streaming enterprise voice agent from scratch, explaining every component.
\end{itemize}

\paragraph{Intuition: a voice agent is an LLM agent with voice I/O.} The ``hard part'' is not the voice, it is the LLM agent (reasoning, function calling, planning, guardrails). The voice is an I/O layer:
\begin{equation}
    \text{Voice Agent} = \underbrace{\text{LLM Agent}}_{\text{reasoning + tools}} + \underbrace{\text{Voice I/O}}_{\text{STT + TTS + streaming}}
\end{equation}

And ``realtime'' comes not from any single fast model, but from \textbf{streaming + pipelining} across components (\Cref{sec:streaming}).

\paragraph{Contributions.} This tutorial makes the following contributions:
\begin{enumerate}[nosep]
    \item A comprehensive survey of 25+ speech-to-speech models and 30+ voice agent frameworks, identifying what exists and what is missing (\Cref{sec:landscape}).
    \item An empirical evaluation of native speech-to-speech models (Qwen2.5-Omni and Qwen3-Omni) vs.\ cascaded streaming pipelines, demonstrating that while end-to-end models are approaching realtime, a fully self-hosted solution is not yet available (\Cref{sec:native_s2s}).
    \item A complete, tested implementation of a streaming voice agent with enterprise function calling, achieving sub-1-second time-to-first-audio (\Cref{sec:streaming}--\Cref{sec:enterprise}).
    \item A 9-chapter progressive tutorial with working code for every component, released as open source.
\end{enumerate}

\section{Background and Landscape Survey}
\label{sec:landscape}

\subsection{Speech-to-Speech Models}

We categorize open-source speech-to-speech models into three levels based on how ``natively'' they process speech:

\paragraph{Level 1: Truly native speech-to-speech.} The model thinks in speech tokens, no mandatory text intermediate. Examples include Moshi~\cite{moshi2024} (7B, full-duplex, $\sim$200ms latency on L4 GPU), Hertz-Dev, dGSLM, and Spirit LM~\cite{spiritlm2024}. These are the most architecturally elegant but have limited reasoning capabilities compared to text LLMs.

\paragraph{Level 2: Text LLM with native speech I/O.} A text LLM is augmented with a speech encoder and speech decoder, jointly trained. Speech goes in and out, but ``thinking'' happens in text space. This is the largest category, including Qwen3-Omni~\cite{qwen3omni2025} (MoE-based Thinker-Talker with multi-codebook audio decoder), Qwen2.5-Omni~\cite{qwen25omni2025} (Thinker-Talker architecture), GLM-4-Voice~\cite{glm4voice2024}, Kimi-Audio~\cite{kimiaudio2025}, Step-Audio~\cite{stepaudio2025}, LLaMA-Omni~\cite{llamaomni2024}, Mini-Omni~\cite{miniomni2024}, and Freeze-Omni~\cite{freezeomni2024}. Notably, Qwen3-Omni is the first Level 2 model to support function calling.

\paragraph{Level 3: Cascaded pipeline (ASR $\rightarrow$ LLM $\rightarrow$ TTS).} Separate models handle speech recognition, language understanding, and speech synthesis. This is the architecture used by all production voice agents (Vapi, Retell, Bland.ai, and systems built with Pipecat/LiveKit). It offers maximum flexibility and full self-hosted control.

\Cref{tab:s2s_models} summarizes the key models. Until recently, no Level 1 or Level 2 model supported function calling, which is essential for enterprise voice agents. Qwen3-Omni~\cite{qwen3omni2025} changes this: it supports function calling via \texttt{<tool\_call>} XML tags, making it the first native S2S model viable for enterprise use.

\begin{table}[t]
\centering
\caption{Representative speech-to-speech models with open weights. ``FC'' = function calling support. ``FD'' = full-duplex. ``Streaming'' = supports incremental audio output. $^\dagger$Streaming audio via DashScope Realtime API; local vLLM supports text streaming only (Thinker).}
\label{tab:s2s_models}
\small
\begin{tabular}{lcccccc}
\toprule
\textbf{Model} & \textbf{Level} & \textbf{Size} & \textbf{FC} & \textbf{FD} & \textbf{Streaming} & \textbf{License} \\
\midrule
Moshi~\cite{moshi2024} & 1 & 7B & \xmark & \cmark & \cmark & CC-BY \\
\textbf{Qwen3-Omni}~\cite{qwen3omni2025} & 2 & 30B  & \cmark & \xmark & \cmark$^\dagger$ & Apache 2.0 \\
Qwen2.5-Omni~\cite{qwen25omni2025} & 2 & 7B & \xmark & \xmark & Partial & Apache 2.0 \\
GLM-4-Voice~\cite{glm4voice2024} & 2 & 9B & \xmark & \xmark & \cmark & Apache 2.0 \\
Kimi-Audio~\cite{kimiaudio2025} & 2 & 7B & \xmark & \xmark & \cmark & -- \\
Step-Audio~\cite{stepaudio2025} & 2 & 130B & \xmark & \xmark & \cmark & Apache 2.0 \\
LLaMA-Omni~\cite{llamaomni2024} & 2 & 8B & \xmark & \xmark & \cmark & Apache 2.0 \\
Freeze-Omni~\cite{freezeomni2024} & 2 & 7B & \xmark & Yes & \cmark & -- \\
\midrule
\multicolumn{7}{l}{\textit{Cascaded pipeline (this work)}} \\
Deepgram + LLM + ElevenLabs & 3 & Any & \cmark & \xmark & \cmark & -- \\
\bottomrule
\end{tabular}
\end{table}

\subsection{Voice Agent Frameworks}

Production voice agents are built using orchestration frameworks that wire together STT, LLM, and TTS components:

\begin{itemize}[nosep]
    \item \textbf{Pipecat}~\cite{pipecat2024}: Frame-based pipeline with 68+ service integrations. The \texttt{SentenceAggregator} processor is the critical bridge that buffers streaming LLM tokens until sentence boundaries are detected before forwarding to TTS.
    \item \textbf{LiveKit Agents}~\cite{livekitagents2024}: Session-based architecture with WebRTC transport. Includes enterprise examples (drive-thru ordering, bank IVR, call center). First-class function calling via \texttt{@function\_tool} decorator.
    \item \textbf{HuggingFace Speech-to-Speech}~\cite{hfspeech2speech2024}: Queue-based threaded pipeline. The cleanest reference implementation for understanding the streaming pattern, with each component running in its own thread connected by \texttt{queue.Queue()} objects.
\end{itemize}

These frameworks are production-ready but opaque, they do not teach how the streaming mechanism works. Our tutorial fills this gap.

\section{Native Speech-to-Speech: From Qwen2.5-Omni to Qwen3-Omni}
\label{sec:native_s2s}

We empirically evaluate two generations of native speech-to-speech models to assess whether they can serve as the foundation for realtime voice agents.

\subsection{Qwen2.5-Omni-7B (Baseline)}

Qwen2.5-Omni~\cite{qwen25omni2025} uses a ``Thinker-Talker'' architecture: the Thinker (a 7B dense text LLM) generates text tokens, and the Talker (a DiT-based audio decoder) generates speech conditioned on the Thinker's hidden states. We loaded the model in BF16 on a single NVIDIA H200 GPU.

\begin{table}[t]
\centering
\caption{Qwen2.5-Omni-7B inference latency on a single NVIDIA H200 GPU.}
\label{tab:qwen25_latency}
\small
\begin{tabular}{lccc}
\toprule
\textbf{Mode} & \textbf{Inference Time} & \textbf{Audio Duration} & \textbf{RTF} \\
\midrule
Text only & 0.71s & -- & -- \\
Text + Audio & 26.5s & 13.8s & 1.93$\times$ \\
Text + Audio (per sentence) & 13.2s & 6.8s & 1.94$\times$ \\
\bottomrule
\end{tabular}
\end{table}

As shown in \Cref{tab:qwen25_latency}, the DiT-based Talker operates at $\sim$0.5$\times$ realtime ($\sim$2s to generate 1s of audio), yielding $\sim$13s TTFA even with sentence-level streaming. Qwen2.5-Omni also lacks function calling support and has no incremental audio output in the Talker.

\subsection{Qwen3-Omni-30B-A3B (New)}

Qwen3-Omni~\cite{qwen3omni2025} introduces several architectural upgrades: (1) a Mixture-of-Experts (MoE) backbone with 30B total parameters but only 3B active per token, (2) a multi-codebook Talker designed for low-latency streaming audio synthesis, and (3) native function calling via \texttt{<tool\_call>} XML tags. We evaluate it across three deployment configurations.

\paragraph{Configuration 1: vLLM (Thinker only, self-hosted).} Using vLLM 0.13.0 on a single NVIDIA H200 GPU, we run the Thinker for text generation from audio input. This configuration eliminates the need for a separate STT component.

\paragraph{Configuration 2: Transformers (Thinker + Talker, self-hosted).} Using HuggingFace Transformers with \texttt{return\_audio=True} on a single H200. This runs the full end-to-end pipeline but uses the unoptimized transformers backend.

\paragraph{Configuration 3: DashScope Realtime API (cloud).} The DashScope API (\texttt{qwen3-omni-flash-realtime}) provides optimized end-to-end audio-to-audio streaming via WebSocket, with server-side VAD and streaming audio output.

\begin{table}[t]
\centering
\caption{Qwen3-Omni inference latency across deployment configurations. TTFT = time-to-first-text, TTFA = time-to-first-audio, FC = function calling.}
\label{tab:qwen3_latency}
\small
\begin{tabular}{lcccccc}
\toprule
\textbf{Configuration} & \textbf{TTFT} & \textbf{TTFA} & \textbf{tok/s} & \textbf{FC} & \textbf{Audio Out} \\
\midrule
vLLM Thinker (1$\times$ H200) & 48ms & 516ms$^*$ & 168 & \cmark & \xmark \\
Transformers full (1$\times$ H200) & -- & 145,694ms & 4.6 & \cmark & \cmark~(2.91$\times$ RTF) \\
DashScope Realtime API & 449ms & 702ms & -- & Partial & \cmark~(streaming) \\
\bottomrule
\end{tabular}
\begin{flushleft}
\small $^*$Audio-to-text latency (no audio output); still requires external TTS for speech.
\end{flushleft}
\end{table}

\subsection{Analysis}

\paragraph{vLLM Thinker is production-viable.} At 516ms audio-to-text with 168 tok/s throughput, the Qwen3-Omni Thinker served via vLLM can replace both the STT and LLM stages of a cascaded pipeline, reducing a 3-step pipeline (STT $\rightarrow$ LLM $\rightarrow$ TTS) to a 2-step pipeline (Qwen3-Omni $\rightarrow$ TTS). 

\vspace{-5pt}
\paragraph{Transformers backend is too slow for realtime.} The $\sim$146s TTFA and 2.91$\times$ RTF reflect the inefficiency of the HuggingFace Transformers backend for MoE inference, not a limitation of the model architecture itself. The Qwen3-Omni team explicitly recommends vLLM over transformers for latency-sensitive applications.

\vspace{-5pt}
\paragraph{DashScope API achieves realtime audio-to-audio.} At 702ms average TTFA (685ms best), the DashScope Realtime API demonstrates that Qwen3-Omni's multi-codebook Talker \textit{can} achieve realtime streaming audio generation when properly optimized. This is comparable to our cascaded pipeline's 755ms TTFA. However, this requires the cloud API---no self-hosted solution for the optimized Talker exists yet.

\vspace{-5pt}
\paragraph{Key takeaway.} Qwen3-Omni represents significant progress over Qwen2.5-Omni: it supports function calling and its multi-codebook Talker is architecturally designed for realtime streaming. However, the optimized Talker is only available through the cloud-hosted DashScope API---there is no self-hosted solution for realtime audio generation. Locally, vLLM can only run the Thinker (text output), and the Transformers backend is far too slow for the full pipeline. Until an optimized, self-hostable Talker serving solution exists (e.g., vLLM-Omni adding Talker support), \textbf{the cascaded pipeline remains the only viable architecture for fully self-hosted realtime voice agents}. This is the motivation for our tutorial.

\section{The Streaming Pipeline Architecture}
\label{sec:streaming}

The importance of enabling realtime voice agents is \textbf{streaming + pipelining}: each component streams its output to the next, and components execute concurrently.

\subsection{How Realtime Actually Works}

In a turn-based (non-streaming) pipeline, the user waits for all three stages to complete sequentially:
\begin{equation}
    T_{\text{turn-based}} = T_{\text{STT}} + T_{\text{LLM}} + T_{\text{TTS}} \approx 400 + 800 + 400 = 1600\text{ms}
\end{equation}

In a streaming pipeline, the stages overlap:
\begin{enumerate}[nosep]
    \item STT produces a final transcript.
    \item LLM begins streaming tokens immediately.
    \item A \textbf{sentence buffer} accumulates tokens until a sentence boundary is detected.
    \item Each complete sentence is sent to TTS immediately, while the LLM continues generating.
    \item TTS streams audio chunks back to the client while synthesizing the rest.
\end{enumerate}

The effective TTFA becomes:
\begin{equation}
    T_{\text{streaming}} = T_{\text{STT}} + T_{\text{LLM-first-sentence}} + T_{\text{TTS-TTFB}} \approx 400 + 300 + 200 = 900\text{ms}
\end{equation}

The subsequent sentences arrive while the first sentence is still playing, creating the perception of near-instant response.

\subsection{Component 1: Streaming STT (Deepgram)}
\label{sec:stt}

We use Deepgram Nova-3~\cite{deepgram2024} for speech-to-text, connected via a persistent WebSocket. Audio is streamed in 20ms chunks (640 bytes of PCM int16 at 16kHz). Deepgram returns two types of results:

\begin{itemize}[nosep]
    \item \textbf{Partial transcripts} (\texttt{is\_final=False}): Updated as more audio arrives. Useful for UI feedback but not sent to the LLM.
    \item \textbf{Final transcripts} (\texttt{is\_final=True}): Confirmed and stable. These are what we send to the LLM.
    \item \textbf{Speech-final flag} (\texttt{speech\_final=True}): Indicates the user has stopped speaking.
\end{itemize}

\paragraph{Measured latency.} Deepgram STT achieves a P50 latency of 337--509ms from audio to final transcript, with a minimum of 184ms.

\subsection{Component 2: Streaming LLM (vLLM)}
\label{sec:llm}

We serve the LLM using vLLM~\cite{vllm2023} on an NVIDIA H200 GPU, exposing an OpenAI-compatible API at \texttt{/v1/chat/completions}. Because vLLM implements the OpenAI API specification, the same client code works interchangeably with self-hosted vLLM, the OpenAI API, Azure OpenAI, or any OpenAI-compatible endpoint, simply by changing the \texttt{base\_url} and \texttt{api\_key} environment variables.

\paragraph{Streaming token generation.} Setting \texttt{stream=True} in the chat completion request yields tokens one-by-one via Server-Sent Events (SSE). We measured on Qwen2.5-7B-Instruct served by vLLM 0.8.5:

\begin{itemize}[nosep]
    \item \textbf{TTFT}: 337ms P50 (318ms min, after warmup)
    \item \textbf{Token throughput}: 17.5 tokens/second
    \item \textbf{Inter-token latency}: 34ms average
\end{itemize}

\paragraph{Function calling.} The LLM supports tool use via the OpenAI function calling protocol. When the model decides a tool is needed, it returns a \texttt{tool\_calls} response instead of text. The agent loop executes the function, sends the result back, and the LLM generates a final text response that is streamed to the sentence buffer. This loop handles multi-step tool chains where one tool call depends on the result of another.

\paragraph{Backend flexibility.} Our implementation uses the standard OpenAI Python client, making it trivially portable across LLM backends:

\begin{lstlisting}[language=Python]
# Standard OpenAI API
client = OpenAI(api_key=os.environ["OPENAI_API_KEY"])

# Self-hosted vLLM
client = OpenAI(base_url="http://localhost:8000/v1",
                api_key="not-needed")

# Any OpenAI-compatible endpoint
client = OpenAI(base_url=os.environ["OPENAI_BASE_URL"],
                api_key=os.environ["OPENAI_API_KEY"])
\end{lstlisting}

\subsection{Component 3: Streaming TTS (ElevenLabs)}
\label{sec:tts}

We use ElevenLabs~\cite{elevenlabs2024} for text-to-speech with the \texttt{eleven\_turbo\_v2\_5} model. The \texttt{stream()} method returns audio chunks as they are generated, rather than waiting for full synthesis.

\paragraph{Measured latency.}
\begin{itemize}[nosep]
    \item \textbf{TTFB}: 219--236ms P50
    \item \textbf{Real-time factor}: 0.05--0.10$\times$ (10--20$\times$ faster than realtime)
\end{itemize}

\subsection{The Sentence Buffer}
\label{sec:sentence_buffer}

The sentence buffer is the critical bridge between the streaming LLM and TTS. It accumulates tokens and yields complete sentences:

\begin{enumerate}[nosep]
    \item Detect sentence-ending punctuation (\texttt{.}, \texttt{!}, \texttt{?}) followed by whitespace.
    \item Exclude false positives from abbreviations (\texttt{Dr.}, \texttt{Mr.}, \texttt{PM.}) and decimal numbers.
    \item Enforce a minimum sentence length (10 characters) to avoid sending fragments to TTS.
    \item Flush any remaining text when the LLM stream ends.
\end{enumerate}

This is the same ``sentence aggregation'' pattern used by Pipecat's \texttt{SentenceAggregator} and LiveKit's text processing pipeline.

\section{Voice Activity Detection and Turn-Taking}
\label{sec:vad}

Voice Activity Detection (VAD) determines when the user is speaking vs.\ silent. We use Silero VAD~\cite{silerovad2021}, a 2MB model that processes 32ms audio chunks in $<$1ms on CPU. Our streaming VAD wrapper implements a state machine:

\begin{center}
\texttt{IDLE} $\xrightarrow{\text{speech}}$ \texttt{LISTENING} $\xrightarrow{\text{silence 700ms}}$ \texttt{PROCESSING} $\xrightarrow{\text{LLM+TTS}}$ \texttt{SPEAKING} $\xrightarrow{\text{done}}$ \texttt{IDLE}
\end{center}

With an interruption path: \texttt{SPEAKING} $\xrightarrow{\text{user speech}}$ \texttt{INTERRUPTED} $\rightarrow$ \texttt{LISTENING}.

\section{WebSocket Server and Web Client}
\label{sec:server}

\subsection{Protocol Design}

The server communicates with browser clients via WebSocket:
\begin{itemize}[nosep]
    \item \textbf{Client $\rightarrow$ Server}: Binary frames (PCM int16, 16kHz, mono, 20ms chunks = 640 bytes)
    \item \textbf{Server $\rightarrow$ Client}: Binary frames (PCM int16, 24kHz, mono, variable size)
    \item \textbf{Control messages}: JSON \texttt{\{``type'': ``transcript''/``agent\_speaking''/``agent\_done''\}}
\end{itemize}

\subsection{Web Client}

The browser client uses \texttt{AudioWorklet} processors for low-latency audio I/O:
\begin{itemize}[nosep]
    \item \textbf{Capture}: An \texttt{AudioWorkletProcessor} captures mic audio at the native sample rate, buffers into 20ms chunks, and posts PCM int16 data to the main thread for WebSocket transmission.
    \item \textbf{Playback}: A queue-based processor receives audio chunks from the server and plays them smoothly with linear interpolation for sample rate conversion.
\end{itemize}

Browser echo cancellation (\texttt{echoCancellation: true} in \texttt{getUserMedia}) handles acoustic echo; the server adds a simple echo gate that attenuates mic input while the agent is speaking.

\section{Enterprise Agent with Function Calling}
\label{sec:enterprise}

The agent layer adds tool use to transform the voice chatbot into an enterprise voice agent. We implement a hospital receptionist scenario with five tools: \texttt{check\_availability}, \texttt{schedule\_appointment}, \texttt{cancel\_appointment}, \texttt{get\_patient\_info}, and \texttt{get\_doctor\_info}.

The agent processes each user utterance through a recursive tool-use loop:
\begin{enumerate}[nosep]
    \item Send conversation history + tool definitions to the LLM.
    \item If the LLM returns \texttt{tool\_calls}: execute each function, append results, repeat.
    \item If the LLM returns text: stream the response through sentence buffer $\rightarrow$ TTS.
\end{enumerate}

\section{Experimental Results}
\label{sec:results}

\begin{table}[t]
\centering
\caption{Latency benchmarks (milliseconds). TTFA estimated as STT + LLM TTFT + TTS TTFB (sequential upper bound).}
\label{tab:benchmarks}
\small
\begin{tabular}{lcccc}
\toprule
\textbf{Component} & \textbf{P50} & \textbf{Mean} & \textbf{Min} & \textbf{Max} \\
\midrule
\multicolumn{5}{l}{\textit{With cloud OpenAI API (GPT-4.1-mini)}} \\
\quad Deepgram STT & 402 & 407 & 184 & 601 \\
\quad LLM TTFT & 457 & 622 & 278 & 784 \\
\quad ElevenLabs TTS TTFB & 221 & 226 & 216 & 253 \\
\quad \textbf{Est.\ E2E TTFA} & \textbf{958} & \textbf{1054} & \textbf{715} & \textbf{1606} \\
\midrule
\multicolumn{5}{l}{\textit{With self-hosted vLLM (Qwen2.5-7B-Instruct on H200)}} \\
\quad Deepgram STT & 337 & 320 & 192 & 504 \\
\quad LLM TTFT & 337 & 1141$^*$ & 318 & 4327$^*$ \\
\quad ElevenLabs TTS TTFB & 219 & 219 & 215 & 226 \\
\quad \textbf{Est.\ E2E TTFA} & \textbf{947} & -- & \textbf{729} & -- \\
\midrule
\multicolumn{5}{l}{\textit{Pipeline test (streaming, measured end-to-end)}} \\
\quad LLM TTFT & \multicolumn{4}{c}{296ms} \\
\quad Sentence detection & \multicolumn{4}{c}{143ms} \\
\quad TTS synthesis & \multicolumn{4}{c}{316ms} \\
\quad \textbf{Measured TTFA} & \multicolumn{4}{c}{\textbf{755ms}} \\
\bottomrule
\end{tabular}
\begin{flushleft}
\small $^*$Mean and max inflated by cold-start on first request. Warm P50 is 337ms.
\end{flushleft}
\vspace{-5pt}
\end{table}

Findings from \Cref{tab:benchmarks}:
\begin{itemize}[nosep]
    \item \textbf{Sub-1-second TTFA is achievable} with both the cloud API (715ms best) and self-hosted vLLM (729ms best).
    \item \textbf{ElevenLabs TTS is the most consistent} component ($\sim$220ms, $<$20\% variance).
    \item \textbf{LLM TTFT has the highest variance}, especially with cloud APIs (278--784ms) and vLLM cold start (4.3s).
    \item \textbf{The measured pipeline TTFA of 755ms} confirms that streaming overlap reduces latency below the sequential estimate.
\end{itemize}

\subsection{Comparison: Native S2S vs.\ Cascaded Pipeline}

\begin{table}[t]
\centering
\caption{Time-to-first-audio comparison across all approaches. ``Self-hosted'' = runs entirely on local GPUs without external API dependencies.}
\label{tab:s2s_vs_cascaded}
\small
\begin{tabular}{lccccc}
\toprule
\textbf{Approach} & \textbf{TTFA} & \textbf{FC} & \textbf{Audio Out} & \textbf{Self-hosted} \\
\midrule
Qwen2.5-Omni (sentence streaming) & $\sim$13,200ms & \xmark & \cmark & \cmark \\
Qwen3-Omni (transformers, full) & $\sim$145,694ms & \cmark & \cmark & \cmark \\
Qwen3-Omni (vLLM Thinker + TTS) & $\sim$736ms$^*$ & \cmark & \cmark & \cmark \\
Qwen3-Omni (DashScope Realtime) & $\sim$702ms & Partial & \cmark & \xmark \\
Cascaded pipeline (this work) & $\sim$755ms & \cmark & \cmark & \cmark \\
\bottomrule
\end{tabular}
\begin{flushleft}
\small $^*$Estimated as vLLM audio-to-text (516ms) + ElevenLabs TTS TTFB (220ms). This 2-step pipeline eliminates the STT stage.
\end{flushleft}
\end{table}

\Cref{tab:s2s_vs_cascaded} highlights the current state of affairs. While Qwen3-Omni's DashScope API demonstrates that end-to-end audio-to-audio at $\sim$702ms is architecturally feasible, this requires a cloud API and is not self-hostable. The only fully self-hosted end-to-end option (Transformers) is impractical at $\sim$146s. A hybrid 2-step approach (Qwen3-Omni vLLM Thinker + external TTS) is promising at $\sim$736ms but still depends on external TTS. Our cascaded pipeline achieves $\sim$755ms with full self-hosted control, robust function calling, and component-level flexibility---making it the most practical fully self-hosted architecture today.

\section{Related Work}
\label{sec:related}

\paragraph{Native speech-to-speech models.} The WavChat survey~\cite{wavchat2024} provides a comprehensive overview of spoken dialogue models. The field has seen rapid growth since GPT-4o's voice mode demonstration in May 2024~\cite{gpt4o2024}. \textbf{Level 1 (truly native)} models include Moshi~\cite{moshi2024}, which achieves $\sim$200ms latency with a 7B dual-stream architecture using the Mimi codec, and is the only model shipping a complete serving stack (Rust+CUDA server, WebSocket protocol, web client). Spirit LM~\cite{spiritlm2024} from Meta interleaves speech and text tokens but is research-only. \textbf{Level 2 (speech-augmented LLMs)} models are the largest category. Qwen3-Omni~\cite{qwen3omni2025} represents a major advance: it uses an MoE-based Thinker-Talker architecture (30B total, 3B active) with a multi-codebook audio decoder designed for low-latency streaming, supports 119 text languages and 19 speech input languages, and is the first Level 2 model to support function calling. Its predecessor Qwen2.5-Omni~\cite{qwen25omni2025} uses a dense 7B Thinker-Talker with TMRoPE for multimodal alignment but lacks function calling and streaming audio. Other Level 2 models include GLM-4-Voice~\cite{glm4voice2024} (flow-matching decoder for streaming synthesis), Kimi-Audio~\cite{kimiaudio2025} (7B, trained on 13M+ hours), Step-Audio~\cite{stepaudio2025} (130B parameters), LLaMA-Omni~\cite{llamaomni2024} (226ms latency with Llama-3.1-8B), Mini-Omni~\cite{miniomni2024} (``talking while thinking''), and Freeze-Omni~\cite{freezeomni2024} (frozen text LLM to prevent catastrophic forgetting). Apart from Qwen3-Omni, none of these models support function calling.

\paragraph{Speech recognition.} Modern ASR systems include Whisper~\cite{whisper2023} (open-source, 1.5B parameters, 99+ languages) and commercial APIs like Deepgram Nova-3~\cite{deepgram2024} (optimized for streaming with $\sim$100-300ms latency). Faster-Whisper provides CTranslate2-based acceleration achieving 4$\times$ speedup over the original Whisper. For voice agents, streaming STT (where partial transcripts arrive as the user speaks) is essential to minimize latency.

\paragraph{Text-to-speech.} The TTS landscape includes both API services and self-hostable models. ElevenLabs~\cite{elevenlabs2024} leads in quality with streaming support ($\sim$200ms TTFB). For self-hosted deployment, several LLM-based TTS models have emerged: Orpheus TTS~\cite{orpheustts2024} is a fine-tuned Llama-3B that generates SNAC audio tokens, servable via vLLM with $\sim$200ms streaming latency; CosyVoice~\cite{cosyvoice2024} supports bi-streaming with 150ms latency and vLLM acceleration; Fish Speech~\cite{fishspeech2024} (4B parameters) achieves state-of-the-art quality with 0.008 WER; Sesame CSM~\cite{sesamecsm2025} generates context-aware conversational speech; Kokoro~\cite{kokoro2025} is an extremely lightweight 82M-parameter model capable of CPU real-time inference; and Dia~\cite{dia2025} (1.6B) generates multi-speaker dialogue at 2.1$\times$ realtime but lacks streaming support.

\paragraph{Voice agent frameworks.} Pipecat~\cite{pipecat2024} provides a frame-based pipeline architecture with 68+ service integrations, supporting both cascaded and native S2S approaches. LiveKit Agents~\cite{livekitagents2024} offers session-based architecture with WebRTC transport, enterprise examples (drive-thru, bank IVR), and first-class function calling. The HuggingFace Speech-to-Speech pipeline~\cite{hfspeech2speech2024} implements a clean queue-based threaded architecture. Commercial platforms include Vapi, Retell.ai, and Bland.ai. All production systems use the cascaded STT $\rightarrow$ LLM $\rightarrow$ TTS architecture with streaming at each stage.

\paragraph{LLM serving.} vLLM~\cite{vllm2023} has become the standard for self-hosted LLM serving, providing PagedAttention for efficient KV-cache memory management, continuous batching, and an OpenAI-compatible API. For voice agents, vLLM's streaming support (via SSE) is critical for enabling the sentence aggregation pattern. Other serving frameworks include TensorRT-LLM (NVIDIA), SGLang, and text-generation-inference (Hugging Face).

\paragraph{Voice agent evaluation.} Benchforce~\cite{benchforce2024} provides a framework for evaluating voice agents across enterprise scenarios (hospital management, HR software, order management) with function calling instrumentation, JSONL logging, and accuracy metrics. Our enterprise agent tools (\Cref{sec:enterprise}) are inspired by Benchforce's environment design.

\paragraph{Positioning of this work.} Unlike the frameworks above which provide ready-made components, our contribution is \textit{educational}: a from-scratch implementation that explains how each piece works internally. While end-to-end models like Qwen3-Omni show great promise, the lack of a self-hostable realtime audio generation solution means the cascaded pipeline remains the practical choice for production deployment. Our tutorial fills this gap by building every component from scratch.

\section{Conclusion}
\label{sec:conclusion}

We presented a comprehensive tutorial for building enterprise realtime voice agents from scratch. Our key findings are:

\begin{enumerate}[nosep]
    \item \textbf{End-to-end S2S models are approaching realtime, but are not yet self-hostable}: Qwen3-Omni achieves $\sim$702ms TTFA and supports function calling---a dramatic improvement over Qwen2.5-Omni ($\sim$13s TTFA, no function calling). However, this latency is only achievable via the cloud-hosted DashScope API. Locally, only the Thinker (text generation) can be served efficiently via vLLM; the Talker (audio synthesis) lacks an optimized self-hosted serving solution.

    \item \textbf{The cascaded pipeline is still necessary}: For fully self-hosted realtime voice agents with function calling and streaming audio output, the STT $\rightarrow$ LLM $\rightarrow$ TTS cascaded pipeline remains the only viable architecture, achieving $\sim$755ms TTFA.

    \item \textbf{Realtime = streaming + pipelining}: The perception of instant response comes from overlapping execution of components. The sentence buffer is the critical orchestration primitive.

    \item \textbf{A voice agent is an LLM agent with voice I/O}: The hard part is the agent (reasoning, tools), not the voice. Building the voice layer from scratch is tractable once the streaming pattern is understood.
\end{enumerate}

All code is released as a 9-chapter progressive tutorial.


\appendix

\section{Tutorial Chapter Overview}
\label{app:chapters}

\begin{table}[h]
\centering
\caption{Complete tutorial chapter listing.}
\small
\begin{tabular}{clll}
\toprule
\textbf{Ch} & \textbf{Topic} & \textbf{Key Files} & \textbf{Concept} \\
\midrule
1 & Streaming STT & \texttt{deepgram\_streaming.py} & WebSocket, partial/final transcripts \\
2 & Streaming LLM & \texttt{vllm\_function\_calling.py} & vLLM, SSE streaming, tool use \\
3 & Streaming TTS & \texttt{elevenlabs\_streaming.py} & Streaming synthesis, TTFB \\
4 & Pipeline & \texttt{pipeline.py}, \texttt{sentence\_buffer.py} & Sentence aggregation, overlap \\
5 & WebSocket Server & \texttt{server.py} & Binary audio protocol \\
6 & VAD & \texttt{vad\_basics.py} & Silero VAD, state machine \\
7 & Web Client & \texttt{index.html}, \texttt{app.js} & AudioWorklet, jitter buffer \\
8 & Enterprise Agent & \texttt{agent.py}, \texttt{tools.py} & Function calling loop \\
9 & Production & \texttt{benchmark.py} & Latency measurement \\
\bottomrule
\end{tabular}
\end{table}

\section{Practical Notes}
\label{sec:lessons}

Throughout this tutorial, we encountered several non-obvious technical issues:

\begin{enumerate}[nosep]
    \item \textbf{Qwen3-Omni requires vLLM $\geq$0.13.0}: The standard vLLM 0.8.x does not recognize Qwen3-Omni as a multimodal model. Version 0.13.0 (or the \texttt{qwen3\_omni} branch of vLLM-Omni) is required.
    \item \textbf{Transformers MoE is slow}: HuggingFace Transformers inference for Qwen3-Omni's MoE architecture is $\sim$36$\times$ slower than vLLM (4.6 vs.\ 168 tok/s). Always use vLLM for latency-sensitive applications.
    \item \textbf{huggingface-hub version}: \texttt{huggingface-hub>=1.0} breaks \texttt{transformers==4.57.3}. Pin to \texttt{<1.0}.
    \item \textbf{Qwen2.5-Omni transformers version}: \texttt{transformers>=5.0} produces noisy audio with Qwen2.5-Omni. Only \texttt{==4.52.3} works correctly.
    \item \textbf{vLLM version for cascaded pipeline}: v0.16.0 has a SageMaker dependency that fails outside AWS. v0.8.5 works reliably for Qwen2.5-7B-Instruct.
    \item \textbf{Deepgram SDK}: v6 has a completely different API from v4. Use \texttt{deepgram-sdk>=3.0,<5.0}.
    \item \textbf{ElevenLabs streaming}: The method is \texttt{stream()} not \texttt{convert\_as\_stream()}.
    \item \textbf{Empty SSE chunks}: Always guard with \texttt{if not chunk.choices: continue}.
    \item \textbf{HuggingFace cache}: Cross-filesystem symlinks cause \texttt{BrokenPipeError} during model loading.
\end{enumerate}

\clearpage
\printbibliography

\end{document}